\documentclass[final,tnotealph]{aipproc}
\layoutstyle{8x11single}
\usepackage{amsmath,amssymb}
\usepackage{hyperref}
\newcommand{\beqdvg}{\begin{equation}}
\newcommand{\eeqdvg}{\end{equation}}
\newcommand{\farcsec}{\hbox{$.\!\!^{\prime\prime}$}}

\newcommand{\andvg}[3]{{\it Astr. Nach.}, {\bf #1}, \rm #2 (#3)}
\newcommand{\ajpdvg}[3]{{\it Am. J. Phys.}, {\bf #1}, \rm #2 (#3)}

\newcommand{\baicdvg}[3]{{\it Bull. Ast. Inst. Czek}, {\bf #1}, \rm #2 (#3)}
\newcommand{\bullastdvg}[3]{{\it Bull. Ast.}, {\bf #1}, \rm #2 (#3)}
\newcommand{\crasdvg}[3]{{\it Comptes Rendus Acad. Sci.}, {\bf #1}, \rm #2 (#3)}

\newcommand{\mnrasdvg}[3]{{\it MNRAS}, {\bf #1}, \rm #2 (#3)}

\newcommand{\physrevdvg}[3]{{\it Phys. Rev.}, {\bf #1}, \rm #2 (#3)}
\newcommand{\procroysocadvg}[3]{{\it Proc. Roy. Soc. Series A}, {\bf #1}, \rm #2 (#3)}
\newcommand{\philtransdvg}[3]{{\it Phil. Trans. Roy. Soc.}, {\bf #1}, \rm #2 (#3)}

\newcommand{\sciamdvg}[3]{{\it Sci. American}, {\bf #1}, \rm #2 (#3)}
\newcommand{\jmpdvg}[3]{{\it J. Math. Phys.}, {\bf #1}, \rm #2 (#3)}

\newcommand{\sovastdvg}[3]{{\it Soviet~Ast.}, {\bf #1}, \rm #2 (#3)}

\newcommand{\naturedvg}[3]{{\it Nature}, {\bf #1}, \rm #2 (#3)}

\newcommand{\sciencedvg}[3]{{\it Science}, {\bf #1}, \rm #2 (#3)}

\newcommand{\sovphysdvg}[3]{{\it Soviet~Phys.}, {\bf #1}, \rm #2 (#3)}
\newcommand{\bjhsdvg}[3]{{\it Brit. J. Hist. Sci.}, {\bf #1}, \rm #2 (#3)}

\begin{document}
%
\title[Conceptual origins of gravitational lensing]{The conceptual origins of gravitational lensing}
%
\author{David Valls-Gabaud$^{1,2}$}{
  address={$^1$ Canada-France-Hawaii Telescope,
             65-1238 Mamalahoa Highway,
	     Kamuela, Hawaii 96743, USA}, 
   altaddress={$^2$ GEPI -- CNRS UMR 8111,
	     Observatoire de Paris, 5 Place Jules Janssen,
             92195 Meudon Cedex, France},
   email={david.valls-gabaud@obspm.fr}%
}

\begin{abstract}
We critically examine the evidence available of the early ideas on the bending 
of light due to a gravitational attraction, which led to the concept of
gravitational lenses, and attempt to present an undistorted 
historical perspective. Contrary to a widespread but baseless claim, Newton was 
not the precursor to the idea, and the first Query in his {\sl Opticks} is totally 
unrelated to this phenomenon. We briefly review the roles of Voltaire, Marat, 
Cavendish, Soldner and Einstein in their attempts to quantify 
the gravitational deflection of light. The first, but unpublished, calculations of 
the lensing effect 
produced by this deflection are found in Einstein's 1912 notebooks, where he derived 
the lensing equation and the formation of images in a gravitational lens.  
The brief 1924 paper by Chwolson which presents, without calculations, the formation 
of double images and rings by a gravitational lens passed mostly unnoticed. 
The unjustly forgotten and true pioneer of the subject is F. Link, who not only 
published the first detailed lensing calculations in 1936, nine months prior to 
Einstein's famous paper in {\sl Science}, but also extended the theory to include the 
effects of finite-size sources and lenses, binary sources, 
and limb darkening that same year. Link correctly predicted that the microlensing 
effect would be easier to observe in crowded fields or in galaxies, as observations 
confirmed five decades later. The calculations made by Link are far more
detailed than those by Tikhov and
Bogorodsky. We discuss briefly some papers of the early 1960s which marked the
renaissance of this theoretical subject prior to the first detection of a gravitational
lens in 1979, and we conclude with the unpublished chapter 
of Petrou's 1981 PhD thesis addressing the microlensing of stars in the 
Magellanic clouds by dark objects in the Galactic halo. 
\end{abstract}
\maketitle

\section{Introduction}
It is remarkable to note that the development of gravitational lensing has been 
marked by a wide shift between the 
detailed theoretical calculations, initiated in the 1930s, and the observations
of the corresponding phenomena, the first of which was discovered in 1979. 
The fast pace at which both observational and theoretical progress has been made 
in this subject reflects
the power of the methods and the advance in astronomical instrumentation. This
makes the discipline mature enough to provide ground for its own history. 
Unfortunately most attempts so far have repeated the same baseless stories (with
the notable exception of \citet{Trimble2001}), sometimes even verbatim, without paying
any attention to the actual historical sources. The history of lensing is
intimately related to the history of the concept of black 
holes \citep{Eisenstaedt1991}
and the observational evidence for the gravitational deflection of light
\citep[{\it e.g.},][]{EarmanGlymour1980,Hentschel1997,Crelinsten2006}.
 The purpose of the present paper
is to set the record straight by critically examining the papers and the context
of their production, so as to present --as far as one can-- an undistorted
account of the evolution of ideas leading to the concept of gravitational lenses. A
more detailed review is presented elsewhere \citep{VallsGabaud2006}.\\

The term {\sl gravitational lens} appears in print for the first time in a rather dismissive 
and negative context :
\begin{quote}
[...] {\sl  it is not permissible to say that the solar gravitational field  acts like a lens, for it has no focal length.} \\
$ \; $ \hspace*{12.0cm} Lodge (1919) \citep{Lodge1919}
\end{quote}
Sir Oliver was of course right, and  the reason is simple : the deflection, produced by the Sun at a radial distance $r$, of a light 
ray coming from infinity and reaching the observer is 
$\alpha(r) \; = \; 1\farcsec75 \; \left(M/M_\odot\right) \; 
\left(R_\odot/r \right) $, 
and is therefore equivalent to a strongly astigmatic lens since the deflection
decreases with increasing distance to the optical axis (compare for
instance to the case of a convergent lens, where the deflection increases
with increasing distance from the axis, so that all emerging rays do
converge to a well-defined and unique focus).

This comment, published as a letter to {\it Nature} in the wake of the discussions
on the results of the famous 1919 eclipse expedition, reflects part of the
huge polemic that arose from the experimental verification of Einstein's theory. 
Although much of it has now faded, given the resounding experimental 
success of General Relativity on all fronts, the issue arose due to the uncertainties 
in the measures at both Sobral and Principe, the two sites of the
expedition to observe the eclipse of May 29, 1919. The plates taken with 4-in telescope at Sobral yielded a weighted average deflection of 1\farcsec98$\pm$0\farcsec18 (after taking into account part of the systematics), while the plates taken with the 13-inch 
astrographic telescope of the Greenwich Royal
Observatory were slightly diffuse, as a consequence of the change of focus
caused by the heating of the mirror of the coelostat, and produced a
measured deflection of 0\farcsec93. The observations at Principe were in
principle of better quality, thanks to the temperature stability of the
island, but were affected by clouds and yielded 1\farcsec61$\pm$0\farcsec30 
\citep{Dyson1920}. The
gravitational deflection was clearly measured, and corresponded to the general
relativistic prediction \citep[see, for instance][for more details.]{EarmanGlymour1980,Hentschel1997,Crelinsten2006} 


It was in the context of this announcement that J.J. Thomson, president  of
the Royal Society at that time, and chair of the joint meeting of the
 Royal Society and the Royal Astronomical
Society on November 6 1919, 
claimed that Newton had thought about the gravitational attraction of light by matter :
\begin{quote}
{\sc The president.} I know call for discussion on this momentous communication. If the results obtained had been only that light was affected by gravitation, it would have been of the greatest importance. Newton did, in fact, suggest
this very point in the first query of his `Optics' and his suggestion would
presumably have led to the half-value. [...]
\end{quote}
as reported in the minutes of the meeting (Royal Astronomical Society, 1919). It is unclear
whether the large portrait of Isaac Newton in the background was
decisive for this statement to be made or if some nationalism was required
 to compensate the intellectual `victory' by a German scientist on the wake of the
end of the first World War.

J.J. Thomson was referring to what was going to become the most famous quote 
extracted from Newton's {\it Opticks} (1704,\citep{Newton1704} Book IV, Part 1), which reads 
\begin{quote}
{\sl Query 1. Do not Bodies act upon Light at a distance, 
and by their action bend its Rays; and is not this action
 ({\it caeteris paribus}) strongest at the least distance?}\hfill  Newton (1704)\citep{Newton1704}
\end{quote}
This is the very first Query that appears in the Book IV, Part 1 of the first
edition of his {\it Opticks} published in 1704. The citation got a widespread
diffusion and is to be found in almost every historical introduction on
gravitational lensing.
Unfortunately the citation is taken totally out of context and has nothing to do
with a gravitational bending of light. 

\section{The fist query of Newton's {\sl Opticks}}
\label{sec:dvg:newton}

The actual context (end of Book III) is the following :
\begin{quote}
{\sl Obs. 10. When the Fringes of the Shadows of the Knives, 
fell perpendicularly upon a Paper at a great distance
 from the Knives, they were in the form of Hyperbola's} 
\end{quote}
that is, Newton is repeating Grimaldi's experiment on {\it diffraction} and
there is no discussion whatsoever on the gravitational effects on light.
As a matter of fact, Newton continues thus :
\begin{quote}
{\sl When I made the foregoing Observations, I design'd to repeat 
 most of them with more care and exactness, and to make new ones
 for determining how the Rays of Light are bent in their passage
 by Bodies, for making Fringes of Colours with the dark lines
 between them. But I was then interrupted} [...]\\

 {\sl And since I have not finish'd this part of my Design, I shall
 conclude with proposing some Queries, in order to a farther
 search to be made by others.}
\end{quote}
This is the end of Book III, Part I. There is no Part II, and this abrupt 
end is very revealing\footnote{In fact, Newton  indicates in his {\sl Advertisement I} 
that his book is composed of various bits and pieces :
\begin{quote}
{\sl Part of the ensuing Discourse about Light was written at the desire of some gentlemen of the Royal
Society, in the year 1675, and then sent to their Secretary, and read at
their meetings, and the rest was added about twelve years after to complete
the theory; except the third Book, and the last proposition of the second, which
were since put together out of scattered Papers.}
\end{quote}
}.
While the first (English) edition of 1704 contained only 16 queries, the
second (1717) has 32, while the Latin edition published in 1706 had 23
\footnote{A rather complete collection of the editions of Newton's works is available on-line at : 
\url{http://dibinst.mit.edu/BURNDY/Collections/Babson/OnlineNewton}}.
The later ones became short essays, in contrast to the brief questions of the
first edition. 
The reason of the ``delay'' in the publication is simple : Robert Hook had
died in 1703.  Hook, along with Pardies, was one proponent of a wave theory of 
light and also a strong opponent of Newton's theory of colour as described
in 1672. The controversies are well documented and will not be reviewed here.
It is important to note that, unlike the {\sl Principia}, the {\sl Opticks} are
first published in vernacular, to maximise the potential number of readers
and to get a quick and widespread diffusion.

The first seven queries will not be changed through the various editions, and
it has to be noted that the rhetorical form of Query is used by Newton to
affirm statements for which he has no proofs. All the questions are formulated to
always give implicitly affirmative answers.

Returning to the subject matter of the first queries, they do indeed all ponder about 
the nature of this bending produced by  diffraction\footnote{The term inflection is
used instead of diffraction, but they are the same concept. For instance, 
Priestley (1772) \citet{Priestley1772} states the following definition in his glossary :
\begin{quote}
{\sl {\it Inflected rays} : those rays of light which, on their near approach to the edges of bodies, in passing by them, are
bent out of their course, being turned either from the body or towards it. This
property of the rays of light is generally termed {\it diffraction} by foreigners, and
Dr. Hooke sometimes called it {\it deflection}.}
\end{quote}
} :
\begin{quote}
\noindent {\sl Query 2. [...]and after what manner are they inflected to make 
those fringes ? \\
Query 3. Are not the Rays of Light in passing by the edges 
 and sides of Bodies, bent several times backwards and forwards [..]\\
Query 4. Do not the Rays of Light which fall upon Bodies 
begin  to bend before they arrive at the Bodies?\\
Query 5. Do not Bodies and Light act mutually upon one another?}
\end{quote}

Newton is of course rightly worried about diffraction,
a phenomenon difficult to explain within his corpuscular theory of light, and 
he is not thinking on the possible effects of gravitation
on the light rays. 
The only mention of the effect of attraction on rays of light can be found
in the {\it Principia} (Scholium to Prop. XCVI in Principia, Book I, Section
XIV), but the attraction there is generic, not necessarily gravitational.

\section{Jean-Paul Marat : a possible precursor}
\label{sec:dvg:marat}

Given the (unfair) black legend that surrounds Marat, it is rather unexpected to
find him as a possible precursor to the idea of gravitational lensing. Marat was
not only the revolutionary politician who played a key r\^ole in the early French
revolution, but also a writer and a physician (he received a MD from St Andrews, 
Scotland in 1775, a year after being admitted as a free mason in London). After 
eleven years spent in England, he returned to Paris in 1776 to become the appointed 
physician at the
house of the Comte d'Artois, the king's brother. His success as medical doctor allows
him to buy and make scientific instruments and carries out experiments on a variety
of subjects mainly on fire and on electrical and optical phenomena. The initial
reception of his books was rather warm, some of them being approved by the Acad\'emie 
Royale des Sciences. The experiments described in his published reports 
appear to be very careful and detailed, 
not better nor worse than the average experiment published in the learned journals of the time,
and so the harsh criticisms made by some\footnote{For instance \citet{Gillispie1980} places
him on the same level as charlatans such as Mesmer.} do not seem to be justified. 
A more balanced and fair account is provided by \citet{Conner1997} and by 
\citet{Bernard1993}.

The reception of Newton's ideas in Europe, dominated by the Cartesianism, was quite cold
at the beginning, if not upright negative \citep[{\it e.g.},][]{Guerlac1981}. It took translations, 
lobbying, polemics (Voltaire played a very important r\^ole with his {\sl \'El\'ements de la
philosophie de Newton}\footnote{For a detailed account of Voltaire's ideas
on the bending of light see \citep{VallsGabaud2006}.}) and the expeditions to Lapland
and Peru to measure the length of the arc of meridians for the gradual acceptance of
newtonian ideas in Europe.  In this context, it should be noted that
Marat was one of the first (and faithful) translators into French of 
Newton's {\sl Opticks} and entertained good relations with Benjamin Franklin, but
very bad ones with Lavoisier (both in his r\^ole as Fermier G\'en\'eral and as
an influential member of the Acad\'emie Royale des Sciences, which in the words
of Marat, was becoming a club of modern charlatants).
Remarkably, in his {\sl D\'ecouvertes sur la lumi\`ere} published in 1780, Marat states that 
\begin{quote}
{\sl Tous les corps connus d\'ecomposent la lumi\`ere en l'attirant [...] 
La sph\`ere d'attraction de la lumi\`ere [...] d\'epend de
la densit\'e superficielle, [...] un facteur d'affinit\'e, [...]
et en raison du carr\'e inverse de la distance}\\
$ \, $ \hspace*{12.0cm} Marat (1780)\citep{Marat1780}
\end{quote}
However, the remaining text is ambiguous on whether this is truly 
a {\sl gravitational} effect or mere attraction by yet another force.
One can't but speculate whether Marat had read the encyclopedic compendium
by Priestley (1772\citep{Priestley1772}) while he was living in England. Certainly 
his opthalmologic
experience led him to think on the theory of vision \citep{DeCock1991}. We also note a number 
of similarities such as the use of a
``solar microscope'' and the description of ``sphere of attraction''. Marat's
influence on the subject was, ironically, far greater within the romantics, thanks
to the translation into German of several of his books. Goethe for instance, cites
Marat's ideas when he develops  his theory of colours.

\section{The first correct calculation: Cavendish}
\label{sec:dvg:cavendish}

Among the many papers and documents by Henry Cavendish found  in the 
collection of the
Duke of Devonshire, four were selected in  1921 for their 
astronomical interest :  on the transit of Venus,
on the precession of the equinoxes, on the influence of tides on the
rotation of the Earth, and one on the bending of light by gravitation. \citet{Dyson1921} 
selected it for obvious reasons\footnote{[...]{\sl the possibility of the bending of a ray of light
by a gravitational field is at present engaging attention, though Cavendish was
working on a corpuscular theory.} Dyson's only comment to the
manuscript is the following note : {\sl [This deflection is half the amount given
by Einstein's law of gravitation]}.}, and comments that the calculation
``{\sl may have been suggested by Query 1 of Newton's Opticks}'', repeating the claim made by J.J. Thomson. 
As far as we know, this is the first  document that explicitly states the influence of
gravitation of light. The manuscript is rather cryptic :
\begin{quote}
To find the bending of a ray of light which passes near the surface
of any body by the attraction of that body. 

Let $s$ be the centre of body and $a$ a point of  surface. 
Let the velocity of body revolving
in a circle at a distance $as$ from the body be equal to the velocity
of light as $1:u$, then will the sine of half bending of the ray be equal 
to $1 / 1 + u^2$.
\end{quote} 

Cavendish of course is well known for his discoveries of hydrogen, of the compound nature
of water and for the measure of the gravitational constant and the density of the Earth, 
among many other subjects \citep{JungnickelMcCormmach1999}. Cavendish was in close contact
with John Michell\footnote{well known for his books on artificial magnets and on
earthquakes  but also for his ideas on the astronomical scale
of magnitudes, on the clustering of stars --including the existence of physical 
binaries-- and on  black holes \citet{Eisenstaedt1991}. He also contributed
extensively to Priestley's (1772) {\sl History and present state of discoveries relating to vision, light and
colours}\citep{Priestley1772}.}  on many problems such as the ``Cavendish'' balance, 
and especially on Michell's idea of weighting stars through the fact that if light
is subject to gravitational attraction, its velocity will be decreased. Therefore
the effect can be detected by the difference in refrangibility and Michell proposed
a device based on a prism to measure the effect, which clearly depends on the mass
of the star. Knowing the mass, and having a mass-luminosity relationship would give
the distance of the stars. 
Cavendish supported the idea, which was
presented at the meeting of the Royal Society on November 27, 1783.

The Cavendish manuscript could  well have been written at this time, around 1783 or 1784,
but it turns out that an examination of the
watermark on the paper shows that it could not have been earlier
than 1804 \citep{JungnickelMcCormmach1999}. The calculation is perhaps then
related either to the diffraction grating experiments carried out
by the astronomer David Rittenhouse in 1787 or to Thomas Young's 1800
famous paper putting forward a wave theory of light. It seems unlikely 
that Cavendish had read Soldner's paper (discussed in the next section) which
deals explicitely with the same problem. Alternatively, Cavendish may have
carried out the calculation reading the discovery of bound binary stars
by Herschel in 1803 and thought again on the Michell effect.
\citet{Will1988} has provided a good way of reaching Cavendish's result, 
which illustrates the proper application of Newtonian dynamics. The important
point to note is that Cavendish takes the proper boundary condition for the
speed of light, which should be taken at infinity, where the acceleration
produced by the Sun's attraction is negligible.

\section{Soldner's mistakes}
\label{sec:dvg:soldner}
%
%
This correct boundary condition is entirely at odds with the result that Soldner, an assistant
of the Prussian royal astronomer J. Bode, published in 1801 \citep{Soldner1801}. Johann
Georg  Soldner was a curious character.  Self-taught, he reaches a prominent 
position at Munich Observatory. He is  well known in geodesy for his 
method for measuring length of arcs over several km with an error smaller than 1 cm, and
has also published some ``cosmological'' works : on motion of stars within the 
Milky Way, and on the invisible (Laplace-Michell) 
stars \citep[see][for a detailed account]{Jaki1978}.

The aims that Soldner sets in his paper are clear from the title :
\begin{quote}
{\sl On the deviation of a light ray from its motion
along a straight line through the attraction of a celestial body
which it passes close by} \\
{\sl [...] to derive all circumstances that exercise
an influence on the true or mean position of a celestial body from
the general properties and interactions of matter}
\end{quote}
He considers the case of a star at the horizon and asks what is the effect, besides
refraction, that will change the apparent position. The attraction produced
by the Earth will clearly bend 
the trajectory and he wants to compute the corresponding astrometric error.
It is a difficult paper to read (but see \citep{Jaki1978} for an English translation),  with a confusing notation where for instance
the acceleration is defined as $g = s / t^2$
and so $v = 2 g t$ instead of $v = g t$ and $g = s / 2 t^2$. 
 In addition the units are not consistent
(for instance, velocities measured in units of length\footnote{This curious habit is still in
use in some
countries, where for instance the speed limits are given in miles or km, 
not miles per hour or km per hour.}). Crucially there are two 
misprints of a factor of two,  and they  fortuitously cancel at the end. The most
important aspect, however, is that it is conceptually {\it wrong }. 
In contrast with Cavendish's calculations, he 
takes the speed of light at the minimum impact parameter, that is, nearest to the Sun,
where Newtonian theory predicts that light will be accelerated and therefore its
speed will be much larger than at infinity.
The mistake that Soldner makes is to assume that the
ray that leaves tangentially the surface of the star will have the same speed
as the ray that, coming from infinity, will reach the star. The trajectories are
almost the same, but they are not the time reversed equivalents since the speed
of light will increase in falling case, and decrease in the emerging one.
However, he does give the first quantitative result : 0\farcsec001 for the Earth and 
0\farcsec84  at the limb of the Sun. He concludes that
\begin{quote}
{\sl Though the combination of several bodies which a light ray could encounter on its way, would be a larger result, for our
observations it is nevertheless unnoticeable. Therefore it is clear that nothing makes
it necessary, at least in the present state of practical astronomy, that one should
take into account the perturbation of light rays by attracting celestial bodies.}
\end{quote}
Soldner's paper would resurface over a century later during the infamous antisemitic 
and antirelativist campaign led by Nobel prize winner Lenard, who
 published excerpts of Soldner's paper in 1924 and accused Einstein of plagiarism.

\section{The second calculation: Einstein (1911)}
\label{sec:dvg:einstein1911}

Although this gravitational bending of light does appear in a few books during
the 19c, the effect was largely forgotten and  resurrected only in 1907 when Einstein,
writing a review article on [special] relativity for the {\sl Jahrbuch der Radioactivit\"at
und Elektronik} edited by J. Stark, indicates (at the very end)  
that the light rays must be bent by the gravitational field. 

Einstein will publish the detailed prediction in 1911 \citep{Einstein1911}, during his tenure at the German 
University of Prague, by simply using  the  equivalence principle to get the 
first-order Cavendish-Soldner result. The crucial point here is that with the
astronomical technology of the time the predicted displacement of 0\farcsec83 should
be observable during total eclipses of the Sun \citep{Einstein1911}. 
Curiously enough, this prediction attracted little interest at the time, and only
a junior astronomer at Berlin, Erwin Freundlich, dedicated much efforts to the
experimental verification. The history of the various attempts, from 1912 on,
 at measuring the predicted bending is a fascinating one 
\citep{Hentschel1997,Crelinsten2006}, which culminates with the 1919 eclipse, and
only faded when the modern radar measures fully confirmed the predicted deflection.

It is in this context that  Einstein thinks about the {\sl  lensing} effects
associated with this bending, while sitting 
in the Prague-Berlin train journeys, when he was visiting Elsa. The analysis
of his notebooks he used at this time show \citep{Renn1997} the first derivation of the lensing equation
and the position of the images, in sketches which are completely correct.  
Puzzingly, he forgets about these calculations {\sl completely}, and will not even remark
than on the very same page where one of his papers appears (in an astronomical
journal, no less, \citep{Einstein1924}) there is a brief report by a Physics
professor at Petrograd speculating on the images produced by a gravitational
lens configuration \citep{Chwolson1924}.

\section{ A wild speculation: fake double stars}
\label{sec:dvg:chwolson}

This  professor at Petrograd was no less than O. Chwolson, famous world-wide
through his encyclopedic {\it Treatise of Physics} \citep{Chwolson1906}, which was
translated from the Russian into German, French and Spanish. Einstein thought
highly of these volumes, and Fermi read them over the summer before his first
year at the University of Pisa. The volume on varying magnetic fields gives
a section on special relativity, where the predicted bending of light is
mentioned as one of the possible tests of the theory. Chwolson would update
the volumes, keeping pace of the tremendous progress made in both theoretical
and experimental physics in the 1920s. While not known through his research
papers, he published in 1924 in the astronomical journal of reference, the Astronomischen
Nachrichten, a short note on the possibility of producing 
{\sl fake double stars} due to the gravitational lensing effect. While he makes no
detailed calculations, he points out the effect and remarks that a {\sl perfect
ring} will be produced if the lens, the source and the observer are colinear \citep{Chwolson1924}.
It is not known what the reaction of the local astronomers was, but in any case he
writes that he cannot say whether these phenomena will actually be observed. The
following paper, on the same page, is by Einstein \citep{Einstein1924}, but there
seems to be no trace of a possible interaction between the two on this matter.

\section{The true pioneer : Franti\v{s}ek Link}
\label{sec:dvg:link}
Eddington had used the analogy between the gravitational deflection and
a refraction effect to explain in simple terms both the 1911 and the 1916
predictions. His 1923 book {\sl Mathematical theory of Relativity} became a best-seller
in academic circles and influenced an astronomer noted by his expertise \citep{Link1933} 
on lunar and solar eclipses, F. Link. Link realises
that the Einstein deflection will produce images, in a fashion
similar to the one considered when using refraction by the Earth's atmosphere 
during lunar eclipses. He computes not only the position of the images, but
also their brightness, and considers both visible and invisible stars as 
possible sources, noting that the amplification produced may, in some cases,
render visible an otherwise faint star. As a noted observer, he is
optimistic about the chances of detection of this effect, especially when 
observing spiral nebulae, where the chances of having close-by approaches along
the line of sight are increased:
\begin{quote}
{\sl such close configurations are obviously rare, except in some
areas of the sky, in particular in spiral nebulae.}\\
$ \, $ \hspace*{12.0cm} F. Link (1936) \citep{Link1936}
\end{quote}
 Settling for a while in Paris (although he
will keep strong links with Prague all his life), he publishes his calculations
in the French journals. Realising the importance of his prediction, he sends
his first report to the {\sl Acad\'emie des Sciences}, not only for a peer review, but
also to ensure a quick diffusion. His paper published in the {\sl Comptes Rendus 
de l'Acad\'emie des Sciences} \citep{Link1936}, appears to have passed totally 
unnoticed, especially in the anglo-saxon litterature. The paper was read on
March 16, 1936, and published along with the papers read during that session. 
It predates, therefore, by nine months, the famous Einstein {\sl Science} paper, which
deals with the same subject \citep{Einstein1936}. Over the 1936 summer, Link 
 computes even more subtle details, such as finite-size effects,
including the limb darkening in stellar atmospheres, and will be
published, in French again, in 1937 \citep{Link1937}.
The contents of this amazing 18-pages-long paper are reviewed in  detail
elsewhere \citep{VallsGabaud2006} but can be summarised as follows :
\begin{enumerate}

\item Introduction. Link argues that the experimental confirmation of general
relativity has been difficult, even though the results of the eclipses
show the predicted effect beyond any doubt. Yet, while more precise measures are
obtained, he proposes to use the photometric implications of the deflection.
He notes the short paper by Chwolson who 
showed the geometrical implications, but failed
to give any theoretical background or the photometric effect. He concludes
insisting that the photometric effect can be measurable in some 
cases.

\item The deflection of the light rays. Here he uses Eddington's book (in his
German edition) to get the basic equations for the total and partial
deflections. The deflection depending upon the ratio 
$K=(M/M_\odot)/(R/R_\odot)$ of mass
to radius, in this formalism, he notes that giant stars will have 
$K$ around 0.06 while for the smallest star known at the time $K=600$.

\item Changes in the intensity. He uses his general
photometric theory which he successfully applied to lunar eclipses in 1933, noting that
the formalism applies independently of the underlying physics of the deflection, and
 that in the case of solar eclipses the effect is negligible but
there should be cases where it is important (citing his 1936 short paper).
The general formalism applies to sources and  lenses of finite angular size.

\item Discussion. Link analises the dependence with impact parameter, noting
that the ratio of intensities may become negative for small impact parameters.
This is due to the fact that it is a ratio of variation of areas, and the
absolute value must be taken. He also notes that the light curve is
symmetrical with respect to the time when the impact parameter reaches
the minimum, and the amplification may be smaller than one at large
distances. 

\item Mutual occultation of two stars. The more realistic case where the
lens may occult a fraction of the source is treated, along with the
required effect from a finite-size source. He deals with both a uniform
disc ({\sl i.e.} no limb darkening) and a linear limb darkening term, and notes
the small effect that darkening produces.

\item Invariance of surface brightness. Here he notes that the invariance
of surface brightness combined with the variation of the intensity
implies that the images will be distorted. He presents the first lensing
diagram ever published  and the
positions of the two images produced by gravitational lensing. The calculation 
yields the same result as given previously in his section 5, 
as a mere application of his general lunar eclipse formalism.

\item Variations of the shape of the occulted star [source]. The positions
of the images are analised in the lens plane, as function of the impact
parameter, for the general case. He notes the formation of arclets
and ``lentils'' (counter-images). For a central occultation ({\sl i.e.} zero
impact parameter) a concentric ring is formed, whose radius is very similar to
 the critical radius.  He derives a simple criterion to check whether the
 shadowing effect due to the finite size of the lens can be neglected. He
carries on noting that the net effect will depend on the relative brightness
of the two stars, and that, in general, there will be spectral changes 
at the same time as the photometric effect takes place. He also speculates
that the oddly-shaped deformed images will not be observed except perhaps
with an interferometric method.

\item Numerical examples. Here he considers three cases

\begin{enumerate}

\item An optical double system (i.e. two unrelated stars, close to each
other along the line of sight). He takes as an examples a 
solar-type lens at 2.6 parsecs and a giant at 25.8 pc and computes the
position and brightness of the images, for selected impact parameters, 
 noting that the maximum amplification will be of about 2 magnitudes.

\item A physical double star. Here he considers the binary system similar to 
Sirius and concludes that the effect is negligible, barely 0.05 magnitudes,
due to the large size of the occulted star. Therefore physical
binaries are not interesting configurations due to the small separation
between the components.

\item Clusters of stars (especially globular). Dense clusters are ideal
places to look for this effect, and he starts speculating that the photometric
amplification could be the reason for which the centres of globular
clusters are so bright. He finds that for two stars belonging to the
globular cluster M13, assumed to be at 30 pc [sic], the amplification 
could reach 5.85 magnitudes, but realises that most occultations will
not be central and that there may be many occultations simultaneously
along the line of sight. He concludes the section arguing that the
spiral nebulae are also good places to look for the effect, which
perhaps helps explaining their increased brightness at their centres.
\end{enumerate}

\item Conclusions. Here is the translation of the conclusions :
\begin{quote}
{\sl The reality of the phenomena we have sketched here depends upon the
validity of the Einstein deflection. It is extremely interesting to
look systematically in all the domains of stellar astronomy for
favourable instances where such events can take place, not only to
constitute another proof of the theory, but also as an explanation
of some brightness variations.}
\end{quote}
\end{enumerate}
The second paper concludes stating that it seems {\sl extremely interesting to look
systematically for such phenomena in all domains of stellar astronomy}
 \citep{Link1937}.
This optimism is in sharp contrast with Einstein's  pessimism on this phenomenon.

\section{Mandl and the 1936 Einstein paper}
\label{sec:dvg:einstein1936}
It is indeed a visit paid by Robert Mandl to Einstein on  April 17, 1936, that 
triggers a new calculation by Einstein on the lensing effect \citep{Renn2001}. 
Mandl thinks that the focusing of
light/radiation may explain : (1) the shape of annular nebulae, 
(2) the origion of cosmic rays (amplification of galaxy's light), and 
(3) the extinction of biological species by bursts due
to stellar eclipses.
 Einstein is not impressed, and more than doubtful, to say the least :
\begin{quote}
{\sl I have come to the conclusion that the phenomenon in question will, after all, not be
observable so that I am no longer in favour of publishing anything
about it.} \hfill Einstein, 18 April 1936
\end{quote}
In spite of this, Mandl will continue over the next months 
to press Einstein on publishing these
calculations, and Einstein will eventually, but very reluctantly, agree, 
sending a brief note to {\sl Science} to, in effect, get rid of this
insistence \citep{Einstein1936}. It is not known whether Mandl knew about
Link's paper, but in any case Einstein will remain very pessimistic about
the lensing effect. Unlike Link's papers, the brief note by Einstein 
attracted the attention of many astronomers, from Russell \citep{Russell1937}
to Zwicky \citep{Zwicky1937a,Zwicky1937b}, and, ironically, the paper would
be hailed as the trigger of the subject, even though Link would continue
publishing further detailed predictions \citep{Link1961,Link1967} and
a whole chapter in his monograph on eclipses \citep{Link1969}. Link died in 1984
and it is not known whether he was aware of the discovery of the first lens
in 1979.
While the papers by Link attracted little notice in the West, Russian astronomers
were well aware of them, and some claimed (undue) priority in 1937 
\citep{Tikhov1937,Tikhov1938}.
A detailed account of these contributions, in particular by Bogorodsky, and their influence, is reported elsewhere
\citep{VallsGabaud2006}.

\section{{\sl Annus mirabilis} : 1963--1964}
\label{sec:dvg:sixties}
It is remarkable that, in spite of this series of pioneering papers,
hardly any attempts were made at refining the predictions or starting
a systematic search for the lensing effect in the 1940s and 1950s. Perhaps the fact that general
relativity remained a rather obscure and technical subject even in the physics
departments during these decades could help explaining this (in retrospect)
puzzling situation :
\begin{quote}
{\sl [...] I might refer to the circumstance that from 1936, when I joined the faculty at the University of
Chicago, to 1961, no courses in General Relativity, not even for one single
quarter, were given at the University. And the University of Chicago
 is not atypical.} \hfill Chandrasekhar (1979) \citep{Chandra1979}
\end{quote} 
However, starting in the late 50s, a new series of papers appear on the
subject \citep{VallsGabaud2006}. Darwin \citep{Darwin1958} made a detailed calculation of the positions and
magnifications, 
but takes angles rather than solid angles and so gets the square root of
the correct result. In the Soviet Union, the calculations by 
Idlis and Gridneva \citep{Idlis1960} may be considered as the precursors of the 
weak lensing idea. The subject remains a highly speculative one, as for instance
Metzner \citep{Metzner1963}, who corrects  Darwin's mistake, concurs on the irrelevance
of the topic:
\begin{quote}
{\sl Both Einstein and Darwin have pointed out that the magnitudes involved makes these results more or less
irrelevant from the point of view of observational astronomy.} \citep{Metzner1963}
\end{quote}
 Klimov \citep{Klimov1963} carries out further calculations with galaxies taken 
as lenses, and marks the beginning of serious and detailed studies :
(1) Liebes \citep{Liebes1964} reviews gravitational lensing on all scales: 
stars of the Milky Way (and their optical depth),
globular clusters, unobservable (dark) stars, non-stellar deflectors,
stars in the Andromeda galaxy, gravitational waves, spikes/flashes, and galaxies.
(2) Refsdal \citep{Refsdal1964} publishes in a well-known astronomical
journal, and notes that {\sl It seems safe to
conclude that passages observable from the Earth occur rather
frequently. The problem is to find where and when the passages
take place [...]}
(3) Zeldovich \citep{Zeldovich1964} studies, for the first time, the 
effects of lensing on cosmological scales.

This series will launch the studies, mostly by astronomers, of further details
of the wide range of phenomena produced by lensing. The process will accelerate
after the discovery of the first lens in 1979.
\section{The forgotten chapter of Petrou's PhD thesis}
\label{sec:dvg:petrou}
It was in a different context that Maria Petrou studied the probability of
lensing by objects in the Milky Way's dark halo on stars of the Large Magellanic
Cloud. Her supervisor though the subject too speculative and argued against its
publication, and so only appears as Chapter VII (Lens effect of halo objects) of
her thesis dealing with  {\sl Dynamical models of spheroidal systems}. Remarkably, she concluded
that  the {\sl [...] expected number of amplified stars is about 20} and that 
 {\sl [...]  the variability of the star will be different from other kinds of
variability because there will be no change in its colour}. These are indeed
the figures and properties that, over a decade later, the microlensing experiments
used to detect a new population of dark compact objects in the Galactic halo.

\section{Conclusions}
The conceptual evolution of gravitational lensing brings to light 
the interplay between published and unpublished calculations, wrong and correct
ones, justly influential and unjustly forgotten papers, perhaps just like any
other sub-discipline of intellectual endeavour. However, we can't but try to
set the record straight and acknowledge that F. Link was the true pioneer of the subject
and a visionary, even though he did not succeed in convincing other astronomers,
let alone be recognized in his leading r\^ole. Similarly, 
we can't but speculate on the reasons that led Einstein, who always seeked the 
observational verification of his theoretical predictions (Brownian motion, 
gravitational deflection and redshift), to be --for once-- extraordinarily pessimistic 
on one of Nature's most spectacular phenomena. 


%
\end{document}